\documentclass{article}

\usepackage{changepage}

\usepackage[utf8x]{inputenc}

\usepackage{textcomp,marvosym}

\usepackage{fixltx2e}

\usepackage{amsmath,amssymb}

\usepackage{cite}

\usepackage{nameref,hyperref}

\usepackage{caption}

\date{}

\usepackage{lastpage,fancyhdr,graphicx}
\usepackage{epstopdf}

\usepackage{color}

\usepackage{listings}

\definecolor{deepblue}{rgb}{0,0,0.5}
\definecolor{deepred}{rgb}{0.6,0,0}
\definecolor{deepgreen}{rgb}{0,0.5,0}

\newcommand\pythonstyle{\lstset{
language=Python,
basicstyle=\scriptsize,
otherkeywords={self},             
keywordstyle=\color{deepblue},
stringstyle=\color{deepgreen},
frame=tb,                         
showstringspaces=false            %
}}

\lstnewenvironment{python}[1][]
{
\pythonstyle
\lstset{#1}
}
{}

\newcommand\pythoninline[1]{{\pythonstyle\lstinline!#1!}}

\begin{document}
\vspace*{0.2in}

\begin{flushleft}
{\Large
\textbf\newline{geoplotlib: a Python Toolbox for Visualizing Geographical Data}
}
\newline
\\
Andrea Cuttone\textsuperscript{1*},
Sune Lehmann\textsuperscript{1,2},
Jakob Eg Larsen\textsuperscript{1}
\\
\bigskip
\textbf{1} DTU Compute, Technical University of Denmark, Kgs. Lyngby, Denmark
\\
\textbf{2} The Niels Bohr Institute, University of Copenhagen, Copenhagen, Denmark
\\
\bigskip

* ancu@dtu.dk

\end{flushleft}

\section*{Abstract}
We introduce geoplotlib, an open-source python toolbox for visualizing geographical data.
geoplotlib supports the development of hardware-accelerated interactive visualizations in pure python, and provides implementations of dot maps, kernel density estimation, spatial graphs, Voronoi tesselation, shapefiles and many more common spatial visualizations.
We describe geoplotlib design, functionalities and use cases.

\section*{Introduction}
Geographical data visualization is a fundamental tool for communicating results related to geospatial analyses, and for generating hypotheses during exploratory data analysis~\cite{tukey1977exploratory,andrienko2006exploratory}.
The constantly increasing availability of geolocated data from social media, mobile devices and spatial databases implies that we need new tools for exploring, mining and visualizing large-scale spatial datasets.

The python programming language~\cite{van1995python} has been gaining attention as a data analysis tool in the scientific community~\cite{pythonscience, oliphant2007python} thanks to the clarity and simplicity of its syntax, and due to an abundance of third-parties libraries e.g.~within many disciplines including scientific computing~\cite{van2011numpy,jones2014scipy}, machine learning~\cite{pedregosa2011scikit}, bayesian modeling~\cite{patil2010pymc}, neuroscience~\cite{peirce2007psychopy}, and bioinformatics~\cite{cock2009biopython}.
Currently, however, there is limited support for geographical visualization.

Here, we introduce geoplotlib, a python toolbox for visualizing geographical data.
geoplotlib provides a simple yet powerful API to generate geographical visualizations on OpenStreetMap~\cite{haklay2008openstreetmap} tiles.
We release geoplotlib as open-source software~\cite{geoplotlib}, accompanied by a rich set of examples and documentation.

In the remainder of this paper, we discuss existing tools for geographical visualization and  document the geoplotlib functionalities in detail, and finally we evaluate the computational performance on a large-scale dataset.

\section*{Related work}
In this section we compare existing tools for visualizing geographical data using python.
We divide the related work into three categories: pure python packages, HTML-based packages and Geographical Information System plug-ins.

\subsection*{Pure-python packages}
The matplotlib~\cite{hunter2007matplotlib} library has become the de-facto standard for data visualization in python and provides a large array of visualization tools including scatter and line plots, surface views, 3D plots, barcharts, and boxplots, but it does not provide any support for visualization on a geographical map by default.

The Basemap~\cite{basemap} and Cartopy ~\cite{Cartopy} packages support multiple geographical projections, and provide several visualizations including point plots, heatmaps, contour plots, and shapefiles.
PySAL~\cite{pysal} is an open-source library of spatial analysis functions written in Python and provides a number of basic plotting tools, mainly for shapefiles.
These libraries however do not allow a user to draw on map tiles, and have limited support for custom visualizations, interactivity, and animation.

\subsection*{HTML-based packages}
There is a very rich ecosystem for data visualization for the web.
A number of frameworks allow users to generate plots and charts: we cite as representative Protoviz~\cite{bostock2009protovis}, d3~\cite{bostock2011d3}, Google Charts~\cite{googlecharts}, sigmajs~\cite{sigmajs}.
There is also a large number of libraries for displaying online tile maps, including Google Maps~\cite{googlemaps}, Bing Maps~\cite{bingmaps}, Leaflet~\cite{leafletjs}, OpenLayers~\cite{openlayers}, ModestMaps~\cite{modestmaps}, PolyMaps~\cite{polymaps}. 

In order to generate a HTML visualization from python code, it is needed to generate the HTML and JavaScript code that maps the data to the graphical elements. 
A number of libraries attempt to automate the conversion, such as Folium~\cite{folium}, Vincent~\cite{vincent} and mplleaflet~\cite{mplleaflet}.
This process however is often complex, error-prone and time consuming. 
The complexity can be even greater if some support for animation or interaction is needed. 
Finally, the JavaScript rendering performance may not be adequate for plotting very large datasets.

\subsection*{Geographical Information System plugins}
Geographic Information Systems (GIS) such as QGIS~\cite{qgis2011quantum}, GrassGIS~\cite{grassgis}, ARCGIS~\cite{arcgis}, MapInfo~\cite{mapinfo} provide very powerful tools for spatial data analysis and visualization.
GIS tools usually provide some support for python scripting, although the availability varies from one to another. 
The main limitation of GIS products is their complexity, requiring a significant amount of training to be used effectively, and as discussed before, the need to export the data from python.

\section*{Overview}

\begin{figure}[h!]
  \centering
    \includegraphics[width=1\textwidth]{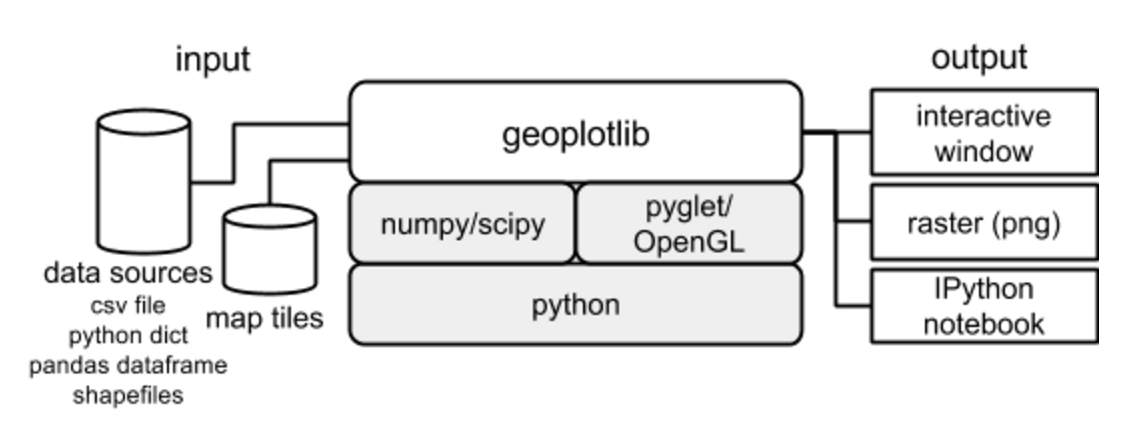}
  \caption{
  Conceptual overview of the geoplotlib architecture.
  geoplotlib builds on top of numpy, scipy, and OpenGL/pyglet.
  It allows to generate geographical visualization as raster, in an interactive window or inside IPython notebooks.
  }
  \label{fig:conceptual}
\end{figure}

An overview of the geoplotlib architecture is given in Fig.~\ref{fig:conceptual}.
geoplotlib builds on top of numpy~\cite{van2011numpy} and scipy~\cite{jones2014scipy} for numerical computations, and OpenGL/pyglet~\cite{pyglet} for graphical rendering. 
geoplotlib implements the map rendering, the geographical projection, the user interface interaction and a number of common geographical visualizations.

\subsection*{Design principles}
geoplotlib is designed according to three key principles:
\begin{itemize}
\item \emph{simplicity}: geoplotlib tries to minimize the complexity of designing visualizations by providing a set of built-in tools for the most common tasks such as density visualization, spatial graphs, and shapefiles.
The geoplotlib API is inspired by the matplotlib~\cite{hunter2007matplotlib} programming model and syntax, the de-facto standard for data visualization in python; this makes it easier for matplotlib users to get started.
\item \emph{integration}: geoplotlib visualizations are standard python scripts, and may contain any arbitrary python code and use any other package. There is no need to export to other formats (e.g. shapefiles, HTML) or use external programs.
This supports a complete integration with the rich python data analysis ecosystem such as scientific computing, machine learning and numerical analysis packages. 
The visualization can even run within an IPython~\cite{perez2007ipython} session, supporting interactive data analysis and facilitating the iterative design for visualizations.
\item \emph{performance}: under the hood, geoplotlib uses numpy/scipy for fast numerical computations, and pyglet/OpenGL for hardware-accelerated graphical rendering.
This allows the visualizations to scale to millions of datapoints in realtime. 
\end{itemize}

\subsection*{A first script}
\label{sec:basic}
A simple geoplotlib script looks like this:

\begin{minipage}{\textwidth}
\begin{python}
data = read_csv('data/bus.csv')
geoplotlib.dot(data)
geoplotlib.show()
\end{python}
\end{minipage}
This script launches the geoplotlib window and shows a dot map of the data points, in this example the location of bus stops in Denmark (Fig. \ref{fig:dot}).
geoplotlib automatically determines the map bounding box, downloads the map tiles, perform the geographical projection, draws the base map and the visualization layers (the dots in this example).
The map is interactive and allows a user to zoom and pan with mouse and keyboard. 

As discussed above, the usage of the geoplotlib API is very similar to matplotlib.
The visualization canvas is initially empty, and each command adds a new layer of graphics.
The geoplotlib window is displayed when \verb|show()| is called.
Alternatively, the map can be rendered to image file using \verb|savefig('filename')|, or displayed inline in an IPython notebook using \verb|inline()|.

\subsection*{Layers}
\label{sec:layers}
The geoplotlib package provides several common geographical visualizations in form of layers.
The API provides convenient methods for quickly adding a new visualization layer.
In this section we provide a summary of the built-in visualizations.
The data for all examples is available on the project website~\cite{geoplotlib}.

\subsubsection*{Dot Map}
\begin{figure}[h!]
  \centering
    \includegraphics[width=1\textwidth]{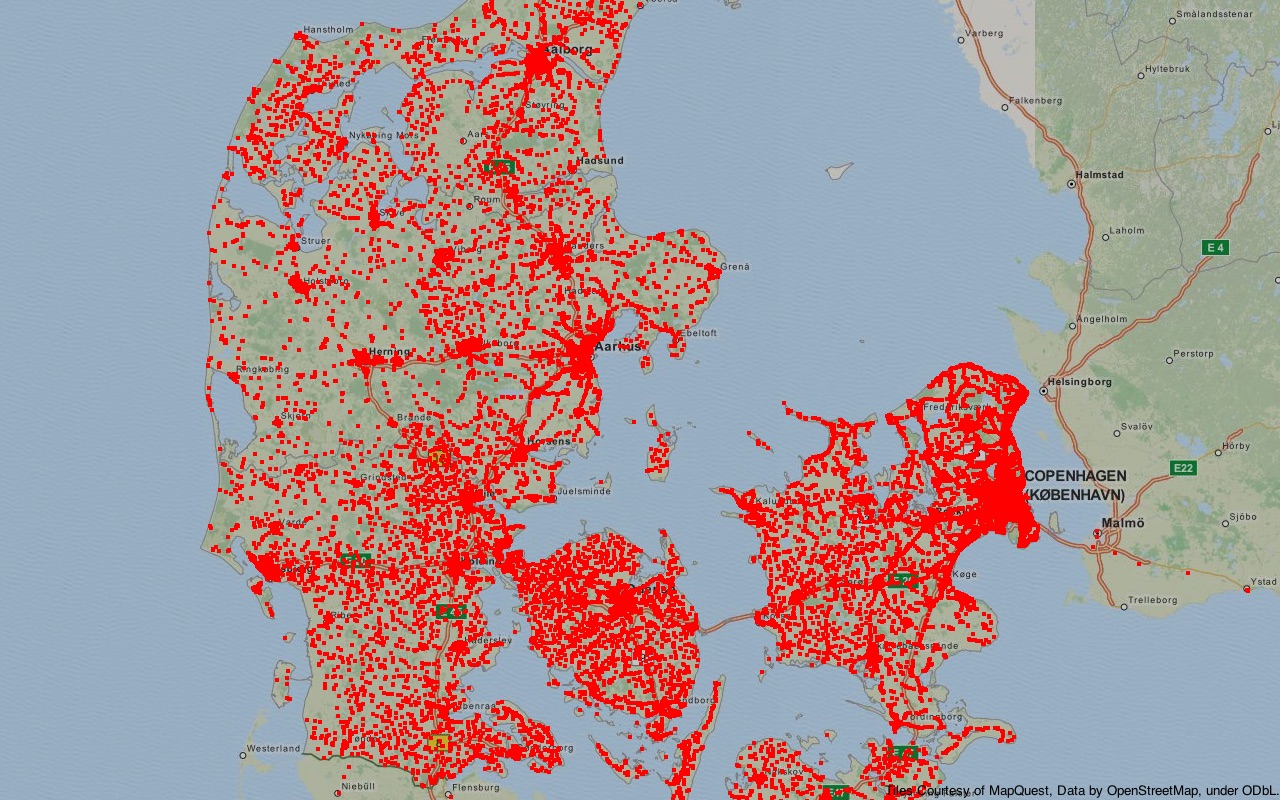}
  \caption{
  A dot map of bus stops in Denmark, where each sample is represented by a point.
  }
  \label{fig:dot}
\end{figure}

An elementary operation in geographical visualization is to display ``what is where'', that is to place a graphic element on the map for each of the objects in consideration.
This provides an immediate idea of the absolute and relative locations of objects.
Moreover, the density of points directly maps to the density of objects on geographical surface, identifying zones of higher and lower density.
An example of dot map is shown in Fig.~\ref{fig:dot}.
The dot map shows the spatial distribution of bus stops in Denmark at a glance.
The zones of higher density -- corresponding to the Copenhagen metropolitan area and to the other major cities are immediately recognizable.
The \verb|dot| method allows users to configure points size, color and transparency, and optionally to attach a dynamic tooltip to each point.

\subsubsection*{2D Histogram}
One limitation of dot maps is that it is hard to distinguish between areas of high density, as the number of point is so high that they uniformly cover the visualization canvas.
A more direct visualization of density is to compute a 2D histogram of point coordinates.
A uniformly spaced grid is placed on the map, and the number of samples within each cell is counted.
This value is an approximation of the density, and can be visualized using a color scale.
In geoplotlib we can generate the 2D histogram of the data using \verb|hist|:

\begin{minipage}{\textwidth}
\begin{python}
data = read_csv('data/opencellid_dk.csv')
geoplotlib.hist(data, colorscale='sqrt', binsize=8)
geoplotlib.show()
\end{python}
\end{minipage}
Here \verb|binsize| refers to the size in pixels of the histogram bins.

\begin{figure}[t!]
  \centering
    \includegraphics[width=1\textwidth]{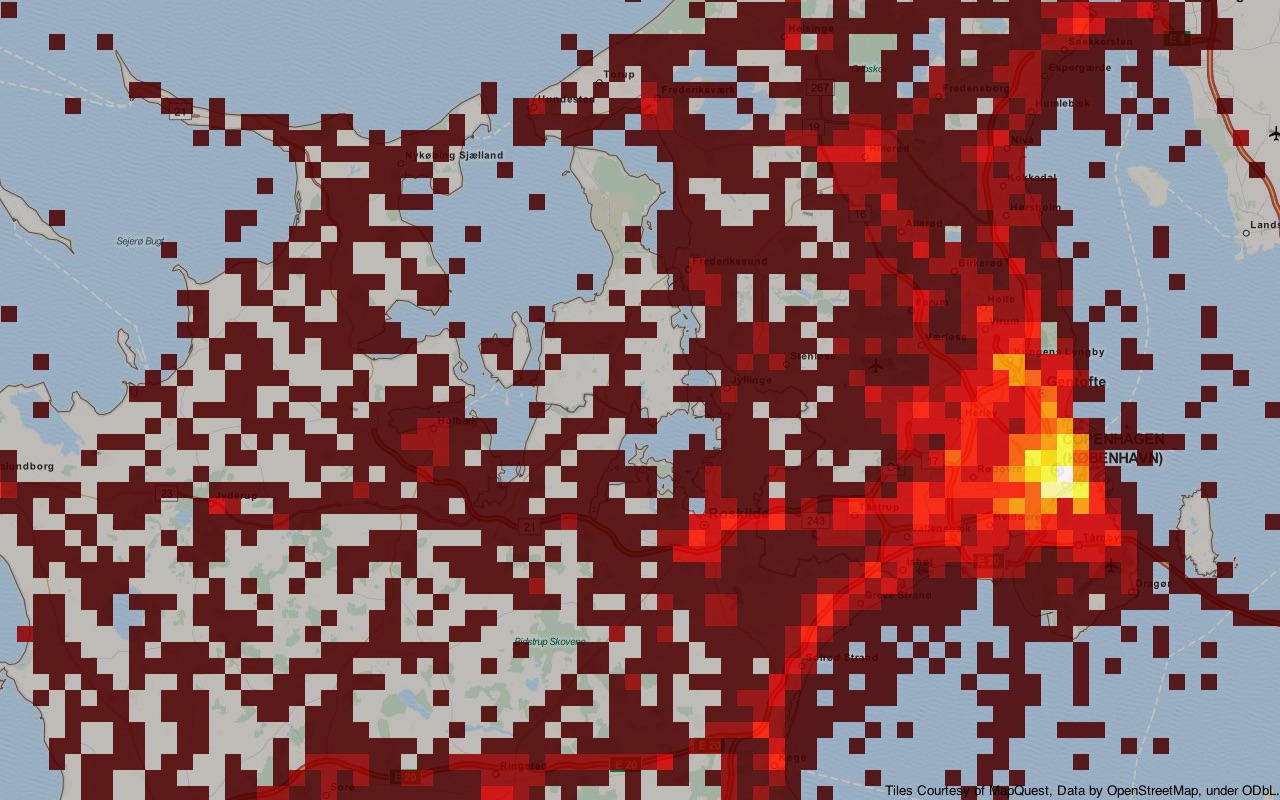}
  \caption{A 2D histogram of cell tower locations in Denmark, using the `hot' colormap (dark red is lower, yellow-white is higher).}
  \label{fig:hist}
\end{figure}

The example above loads some data related to cell tower positions in Denmark, and then generates a histogram with a specific colorscale and bin size (Fig. \ref{fig:hist}).
Compared to the dot map example, the histogram provides a clearer depiction of the density distribution.

\subsubsection*{Heatmap}
\begin{figure}[t!]
  \centering
    \includegraphics[width=1\textwidth]{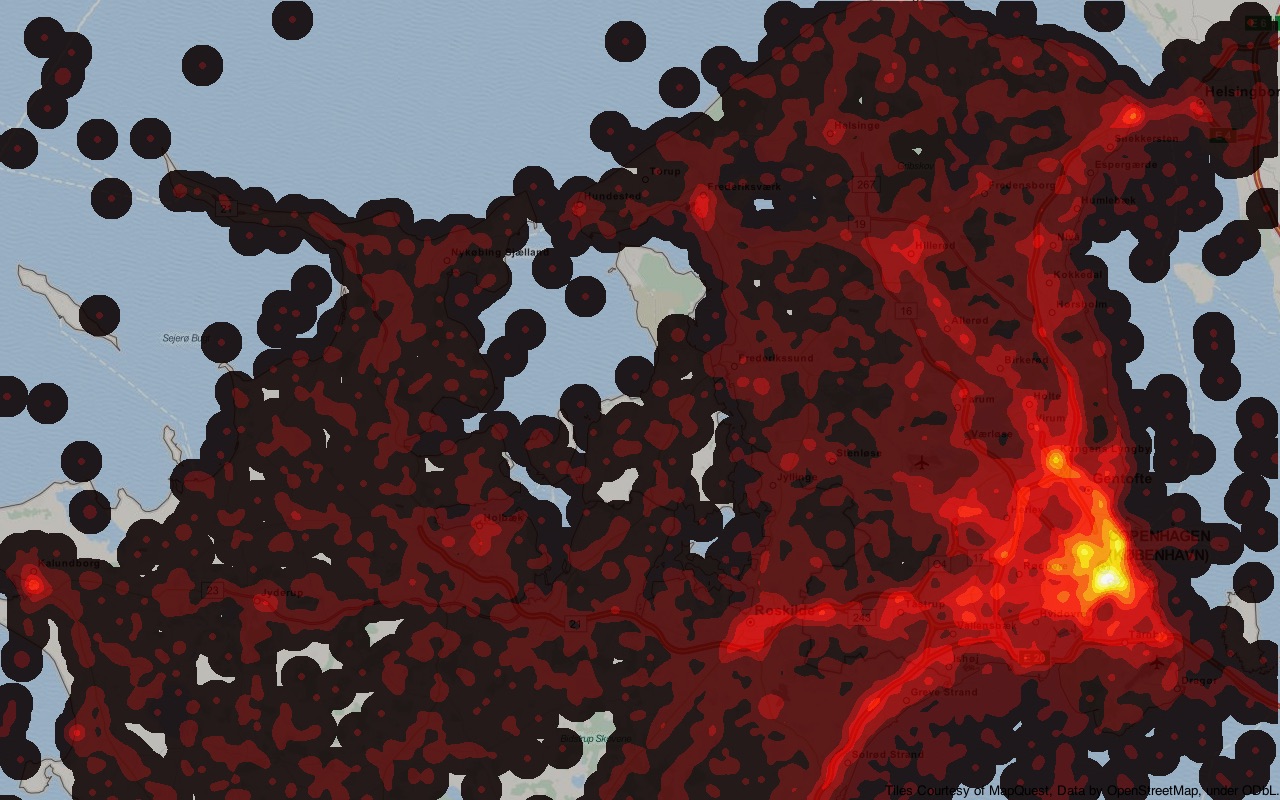}
  \caption{A heatmap (kernel density estimation) of the cell tower locations in Denmark, using a hot colormap (dark red is lower, yellow-white is higher). The kernel density produces a much smoother estimation and therefore a clearer visual representation of the density, if compared with a histogram (Fig.~\ref{fig:hist}).}
  \label{fig:kde}
\end{figure}

The main deficiency of histogram visualizations is that they are discrete approximations of a (effectively continuous) density function. 
This creates a dependence on the bin size and offset, rendering histograms sensitive to noise and outliers.
To generate a smoother approximation, a kernel density estimator approximates the true density function applying kernel functions in a window around each point~\cite{parzen1962estimation}.
The size of this window depends on the bandwidth parameter: a smaller bandwidth will produce more detailed but also noisier estimation, while a larger bandwidth will produce a less detailed but smoother estimation.
A kernel estimation function can then be visualized by a surface where the color encodes the density value (this visualization is often called a ``heatmap'').
In geoplotlib, the \verb|kde| method generates a kernel density estimation visualization:

\begin{minipage}{\textwidth}
\begin{python}
data = read_csv('data/opencellid_dk.csv')
geoplotlib.kde(data, bw=[5,5])
geoplotlib.show()
\end{python}
\end{minipage}
Fig. \ref{fig:kde} shows the kernel density estimation applied to the cell tower data. 
Comparing the histogram from Fig. \ref{fig:hist} with the kernel density estimation in Fig.~\ref{fig:kde}, it is evident how the latter produces a smoother and consequently clearer visualization of density.
The kernel bandwidth (in screen coordinates) can be configured to regulate the smoothness.
The density upper bound can be set to clip density values over a threshold.
Also the density lower bound can be set, to avoid rendering areas of very low density:

\begin{minipage}{\textwidth}
\begin{python}
# lowering clip_above changes 
# the max value in the color scale
geoplotlib.kde(data, bw=[5,5], cut_below=1e-6, clip_above=1)

# different bandwidths
geoplotlib.kde(data, bw=[20,20], cmap='coolwarm', cut_below=1e-6)
geoplotlib.kde(data, bw=[2,2], cmap='coolwarm', cut_below=1e-6)

# linear colorscale
geoplotlib.kde(data, bw=[5,5], cmap='jet', 
            cut_below=1e-6, scaling='lin')
\end{python}
\end{minipage}

\subsubsection*{Markers}
\begin{figure}[t!]
  \centering
    \includegraphics[width=1\textwidth]{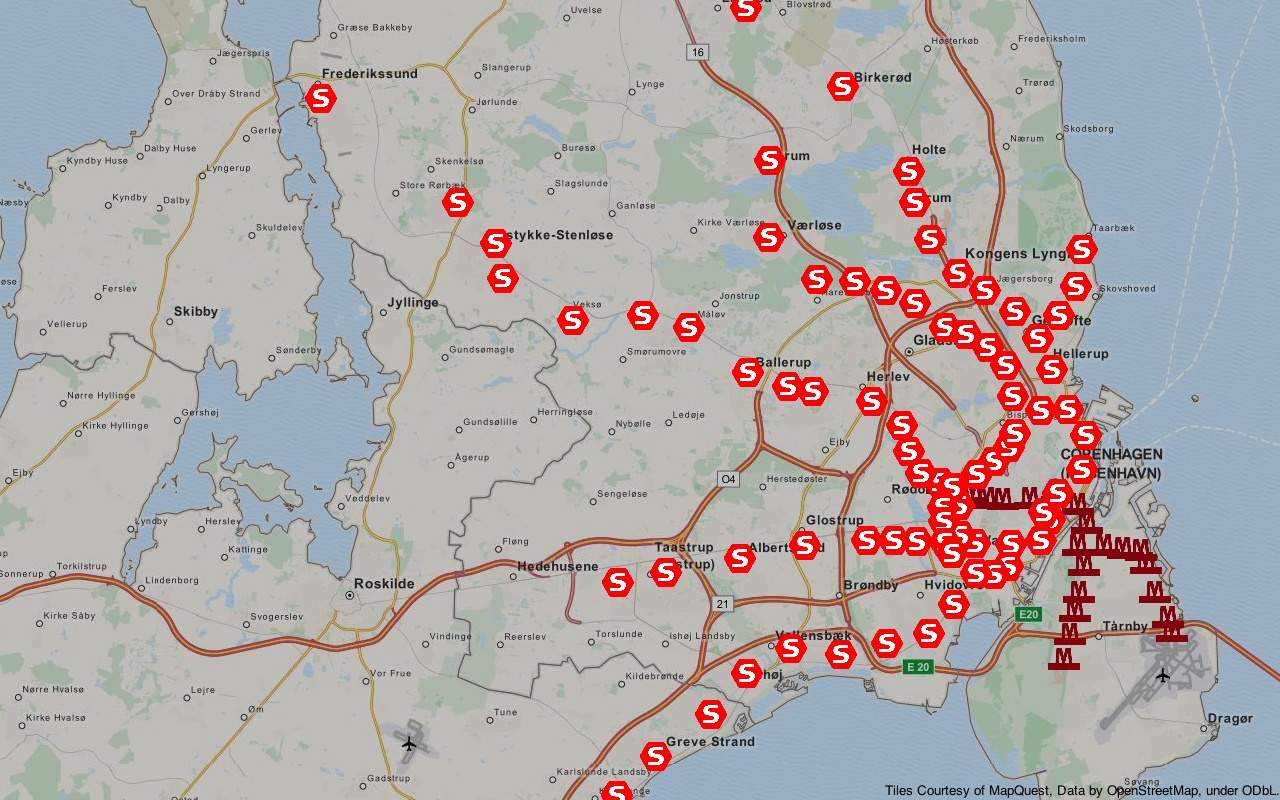}
  \caption{
  Markers showing the location of train and metro stations in the Copenhagen area.
  geoplotlib can use any raster image (png, jpg) as markers.
  }
  \label{fig:markers}
\end{figure}

In some cases it is useful to represent objects on the map using custom symbols with specific meaning.
The \verb|markers| method allows a user to place customs markers on the map:

\begin{minipage}{\textwidth}
\begin{python}
metro = read_csv('./data/metro.csv')
s_tog = read_csv('./data/s-tog.csv')

geoplotlib.markers(metro, 'data/m.png', 
            f_tooltip=lambda r: r['name'])
geoplotlib.markers(s_tog, 'data/s-tog.png', 
            f_tooltip=lambda r: r['name'])
geoplotlib.show()
\end{python}
\end{minipage}
Fig. \ref{fig:markers} shows an example of custom markers for metro and train stops in Copenhagen.
Markers graphics can be any common raster format (png, jpeg, tiff), and can be rescaled to a custom size.
Optionally a dynamic tooltip can be attached to each marker.

\subsubsection*{Spatial Graph}
\begin{figure}[h!]
  \centering
    \includegraphics[width=1\textwidth]{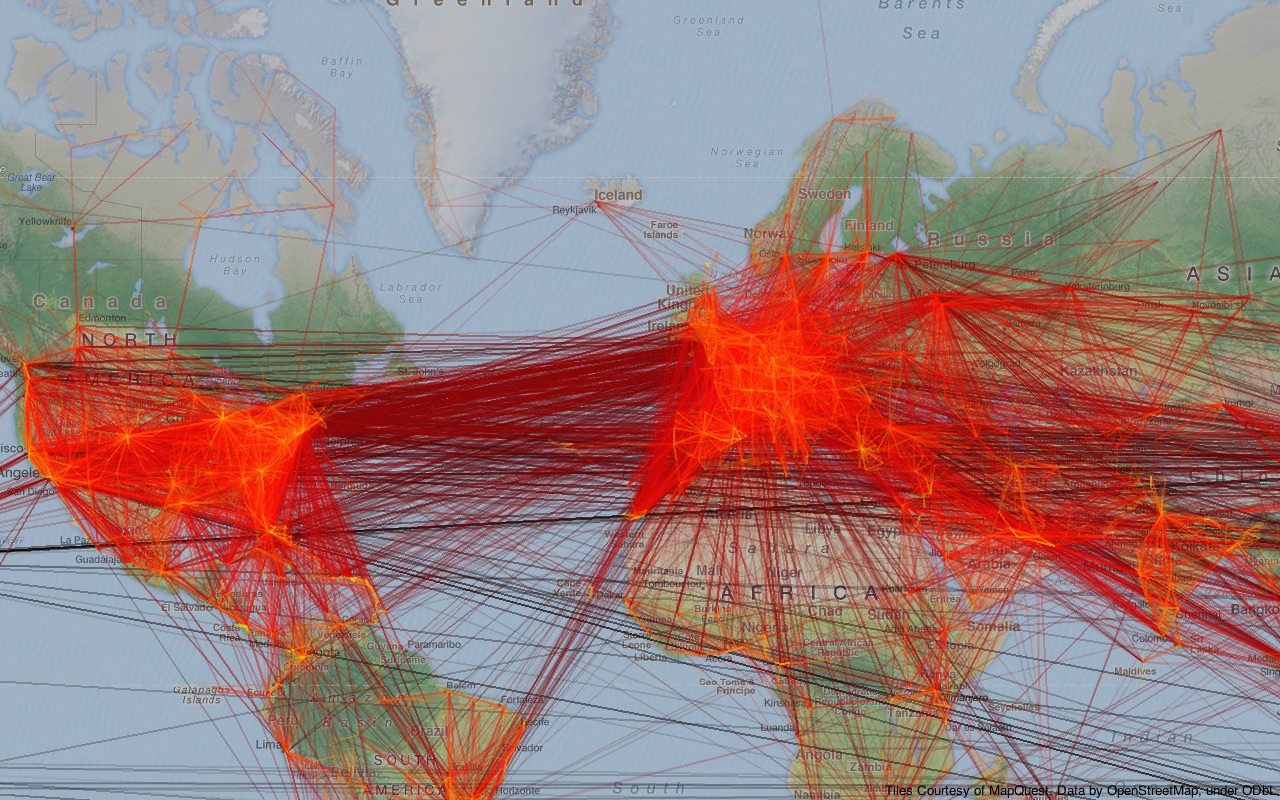}
  \caption{Spatial graph of airport locations, where each node represents an airport and each edge represent a flight connection. Edges are colored using a colormap encoding the edge length.}
  \label{fig:graph}
\end{figure}

Spatial graphs are a special type of graphs where nodes have a well-defined spatial configuration.
Examples includes transport networks (bus routes, train tracks, flight paths), supply chain networks, phone call networks and commute networks.
In geoplotlib \verb|graph| renders a spatial graph:

\begin{minipage}{\textwidth}
\begin{python}
data = read_csv('./data/flights.csv')
geoplotlib.graph(data,
                 src_lat='lat_departure',
                 src_lon='lon_departure',
                 dest_lat='lat_arrival',
                 dest_lon='lon_arrival',
                 color='hot_r',
                 alpha=16,
                 linewidth=2)
geoplotlib.show()
\end{python}
\end{minipage}
Fig.~\ref{fig:graph} shows the resulting spatial graph of airport locations, where each node represents an airport and each edge represents a flight connection. 
Edges are colored using a colormap encoding the edge length.

\subsubsection*{Voronoi Tessellation}
A Voronoi tessellation~\cite{aurenhammer1991voronoi} is a partition of space into regions induced by some seed points, so that each region (called a Voronoi cell) consists of all points closer to a specific seed than to any others.
The analysis of Voronoi tessellation is used in numerous fields including ecology, hydrology, epidemiology, mining and mobility studies.

In geoplotlib \verb|voronoi| can be used to generate a Voronoi tessellation visualization.
Voronoi cell fill, shading and colors can be configured.

\begin{figure}[b!]
  \centering
    \includegraphics[width=1\textwidth]{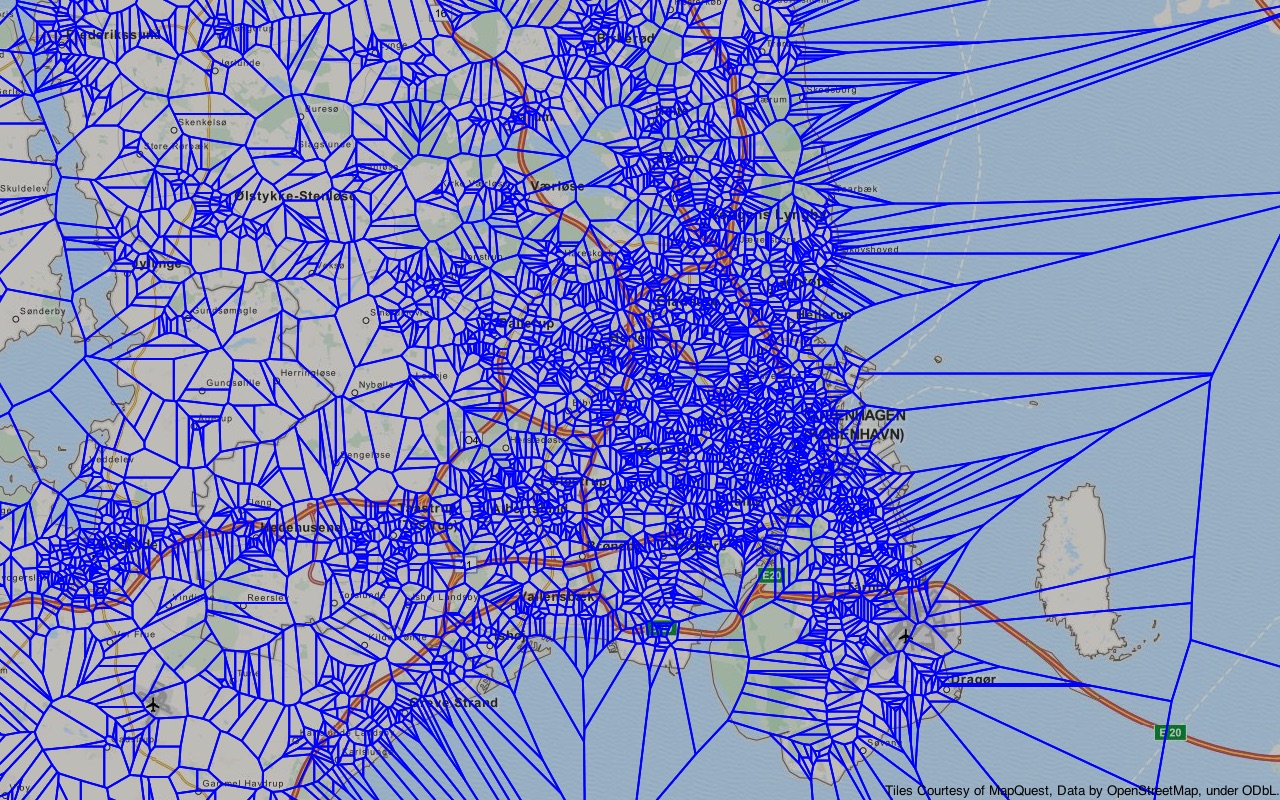}
  \caption{Voronoi tessellation of bus stops in Denmark. Voronoi cells provide an estimation of the space closer to one stop than any others. The density of points is also captured by the size of Voronoi cells, as smaller cells indicate more densely covered areas.}
  \label{fig:voronoi}
\end{figure}

\begin{minipage}{\textwidth}
\begin{python}
data = read_csv('data/bus.csv')
geoplotlib.voronoi(data, line_color='b')
geoplotlib.show()
\end{python}
\end{minipage}
Fig.~\ref{fig:voronoi} provides an example of Voronoi tessellation of bus stops in Denmark. 
Voronoi cells provide a measure of the space closer to one stop than any others. 
The density of points is also captured by the size of Voronoi cells, as smaller cells indicate more densely covered areas.

\subsubsection*{Delaunay triangulation}
A Delaunay triangulation~\cite{de2000computational} is a convenient method for generating triangles meshes from a set of points.
In geoplotlib the \verb|delaunay| method can be used for this purpose.
The edge color can be configured to a fixed value, or to encode the length of the edges.

\begin{figure}[h!]
  \centering
    \includegraphics[width=1\textwidth]{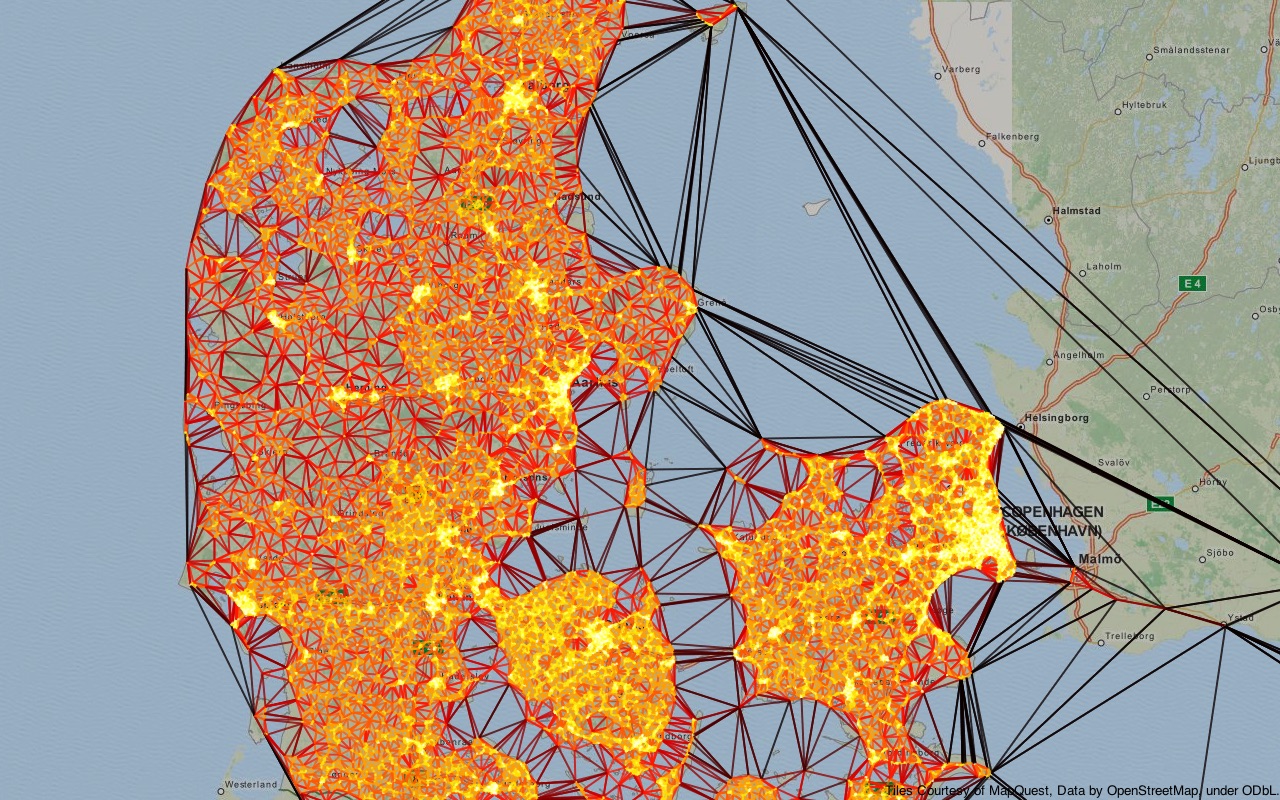}
  \caption{A Delaunay triangulation of the bus stops in Denmark, with edges colored according to length}
  \label{fig:delaunay}
\end{figure}

\begin{minipage}{\textwidth}
\begin{python}
data = read_csv('data/bus.csv')
geoplotlib.delaunay(data, cmap='hot_r')
geoplotlib.show()
\end{python}
\end{minipage}
Fig.~\ref{fig:delaunay} shows the Delaunay triangulation of bus stops, with edges colored according to length.

\subsubsection*{Convex Hull}
\begin{figure}[b!]
  \centering
    \includegraphics[width=1\textwidth]{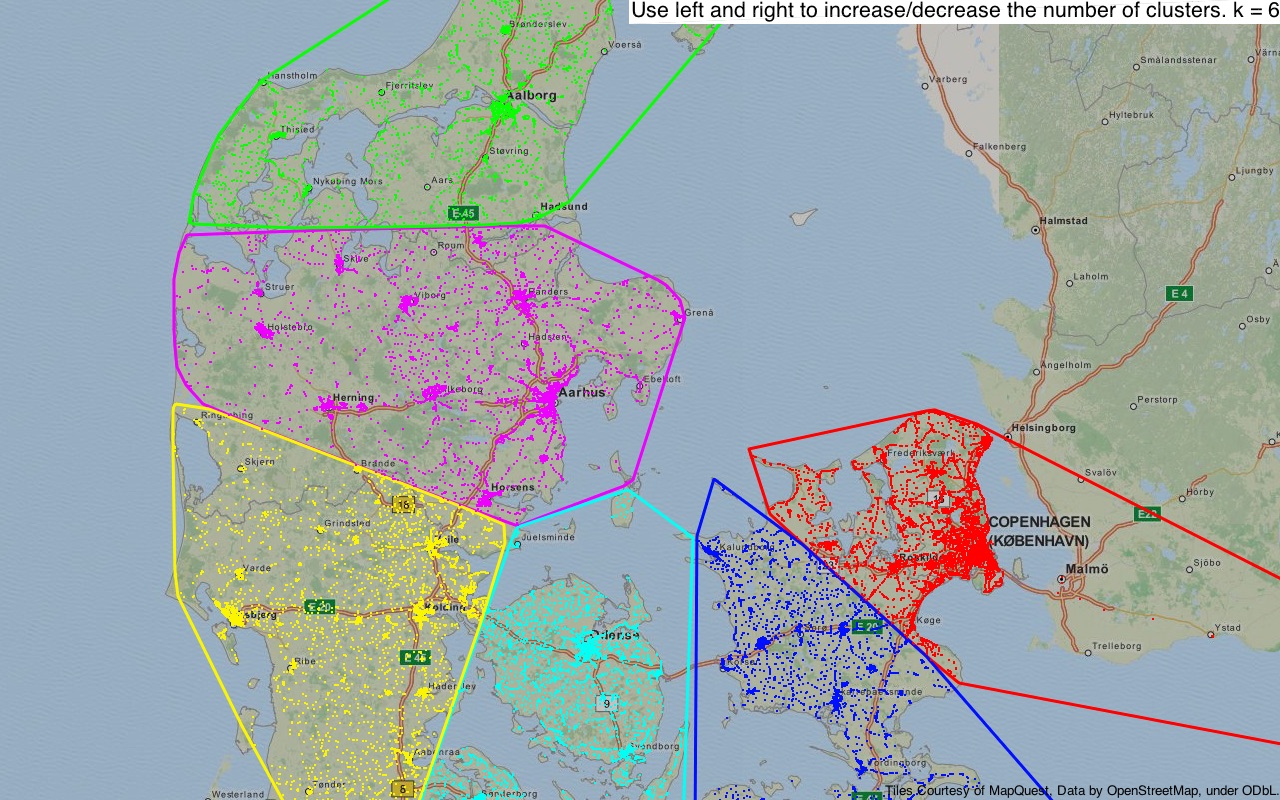}
  \caption{The bus stops points are split into 6 groups, and each group is represented by a different colored convex hull.}
  \label{fig:convexhull}
\end{figure}

A convex hull~\cite{de2000computational} of a set of finite points is the smallest convex polygon that contains all the points.
Convex hulls can be used for example to visualize the approximate area corresponding to a set of points.
In geoplotlib:

\begin{minipage}{\textwidth}
\begin{python}
geoplotlib.convexhull(data, color, fill=True)
\end{python}
\end{minipage}
Fig.~\ref{fig:convexhull} shows the bus stops points split into 6 groups, and each group is represented by a differently colored convex hull.

\subsubsection*{Shapefiles}
\begin{figure}[h!]
  \centering
    \includegraphics[width=1\textwidth]{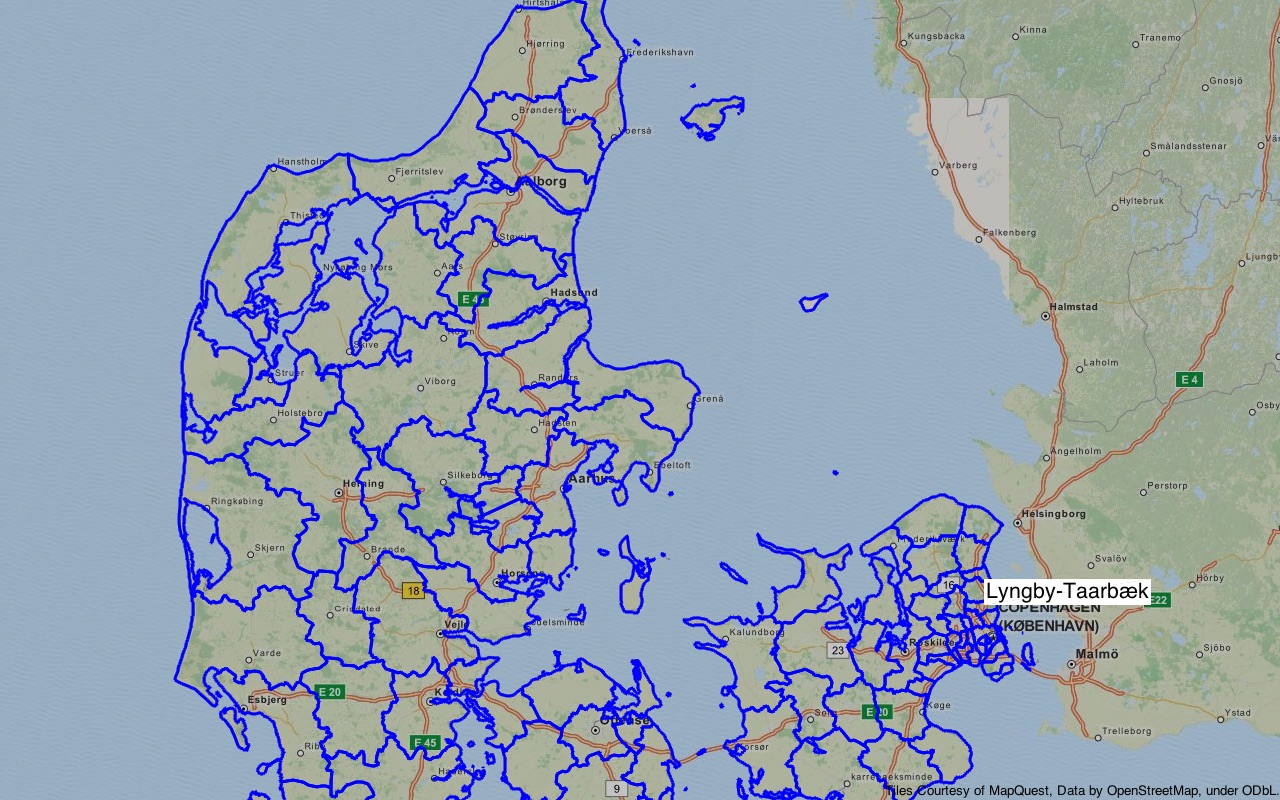}
  \caption{Rendering the shapefiles for the \emph{kommuner} administrative regions in Denmark.}
  \label{fig:shapefiles}
\end{figure}

Shapefile~\cite{esri1998esri} is a popular file format for describing vector graphics for geographical information systems.
geoplotlib uses pyshp~\cite{pyshp} to parse the shapefiles.
The line color can be configured and an optional tooltip can be attached to each shape.
In the following example we display the \emph{kommuner} administrative regions in Denmark (Fig.~\ref{fig:shapefiles}):

\begin{minipage}{\textwidth}
\begin{python}
geoplotlib.shapefiles('data/dk_kommune/dk_kommune',
                  f_tooltip=lambda attr: attr['STEDNAVN'],
                  color=[0,0,255])
geoplotlib.show()
\end{python}
\end{minipage}

\subsubsection*{GeoJSON}
\begin{figure}[b!]
  \centering
    \includegraphics[width=1\textwidth]{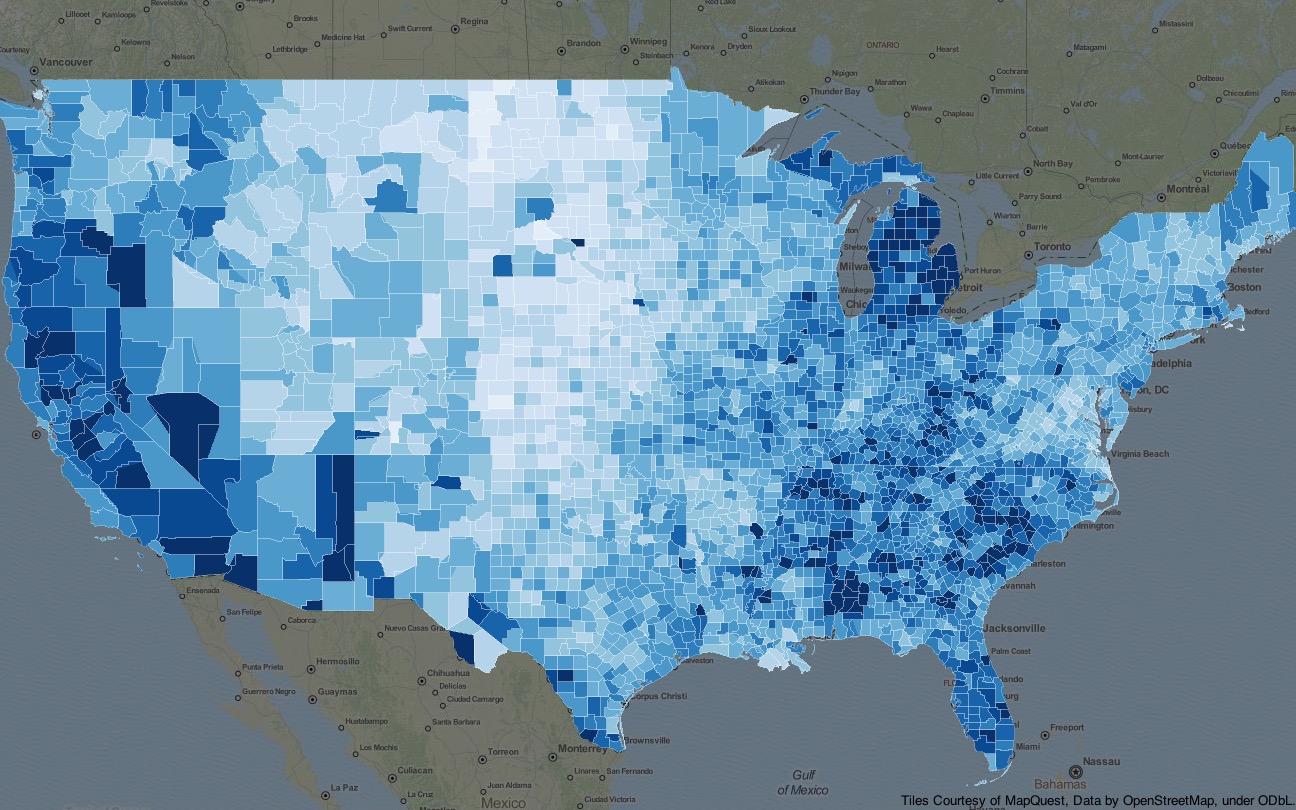}
  \caption{Choropleth of unemployment in USA using GeoJSON shapefiles}
  \label{fig:choropleth}
\end{figure}

GeoJSON~\cite{geoJSON} is a human-readable format for encoding geographical data, such as polygons and lines.
geoplotlib can render shapes from the GeoJSON format, and shape color and tooltip can be dynamically altered to encode data.
For instance GeoJSON shapes can be used to generate a choropleth where each geographic unit is colored to encode a continuous variable.
In the following example (Fig.~\ref{fig:choropleth}) we generate a choropleth of unemployment in USA~\cite{choropleth}:

\begin{minipage}{\textwidth}
\begin{python}
def get_color(properties):
    key = str(int(properties['STATE']))
    key += properties['COUNTY']
    if key in unemployment:
        return cmap.to_color(unemployment.get(key), 
                             .15, 'lin')
    else:
        return [0, 0, 0, 0]

with open('data/unemployment.json') as fin:
    unemployment = json.load(fin)

cmap = ColorMap('Blues', alpha=255, levels=10)
geoplotlib.geojson('data/gz_2010_us_050_00_20m.json', 
    fill=True, color=get_color, 
    f_tooltip=lambda properties: properties['NAME'])
geoplotlib.geojson('data/gz_2010_us_050_00_20m.json', 
    fill=False, color=[255, 255, 255, 64])
geoplotlib.show()
\end{python}
\end{minipage}

\section*{Advanced Functionalities}

\subsection*{Data access}
The \verb|DataAccessObject| class is the fundamental interface between the raw data and all the geoplotlib visualizations. A \verb|DataAccessObject| is conceptually similar to a table with one column for each field and one row for each sample. This paradigm is very common in data analysis terminology, and is equivalent to ndarrays in numpy, and dataframes in pandas and R. A \verb|DataAccessObject| can be initialized by reading a comma-separated values (CSV) file with the built-in \verb|read_csv| method, or can be constructed from a python dict, or from a pandas~\cite{mckinney-proc-scipy-2010} dataframe:

\begin{minipage}{\textwidth}
\begin{python}
dao1 = DataAccessObject({'field1': somevalues, 
                         'field2': othervalues})
dao2 = DataAccessObject(mydataframe)
dao3 = read_csv('somefile.csv')
\end{python}
\end{minipage}
The only two fields required are \verb|lat| and \verb|lon|, which represent to the geographic coordinates. Most of the built-in visualization implicitly refer to these two fields to locate entities in space.
\verb|DataAccessObject| also provides a few method for basic data wrangling, such as filtering, grouping, renaming and deleting rows and columns.

\subsection*{Tile providers}
Any OpenStreetMap tile server can be configured using the \verb|tile_provider| method (users are kindly asked to check the tile usage policy for the selected server, and make sure to provide attribution as needed). 
A number of common free tiles providers are supported, including Stamen Watercolor and Toner~\cite{stamen}, CartoDB Positron and DarkMatter~\cite{cartodb}.

\subsection*{Defining custom layers}
The built-in visualizations provide various commonly used tools for geographical data visualization. 
Multiple layers can be combined into a single visualization for richer display.
For even more complex visualizations, geoplotlib allows users to define custom layers. 
In order to generate a new visualization, a new class extending \verb|BaseLayer| must be defined. 
The custom layer must at least define an \verb|invalidate| and a \verb|draw| method. 
The \verb|invalidate| method is called each time the map projection must be recalculated, which typically happens each time that the map zoom-level changes. 
The \verb|invalidate| method receives a \verb|Projection| object, which provides methods for transforming the data points from the geographic coordinates to screen coordinates. 
The screen coordinates can then be passed to a \verb|BatchPainter| object for the rendering. 
A \verb|BatchPainter| can efficiently draw OpenGL primitives such as points, lines and polygons.
The \verb|draw| method is called at each frame, and typically calls the \verb|batch_draw| method of the painter prepared during \verb|invalidate|.
The following is a complete example of a custom layer, which simply draws samples as points:

\begin{minipage}{\textwidth}
\begin{python}
class CustomLayer(BaseLayer):

  def __init__(self, data):
    self.data = data

  def invalidate(self, proj):
    x, y = proj.lonlat_to_screen(self.data['lon'], 
                                 self.data['lat'])
    self.painter = BatchPainter()
    self.painter.points(x, y)

  def draw(self, proj, mouse_x, mouse_y, ui_manager):
    self.painter.batch_draw()
\end{python}
\end{minipage}
The final step needed is to add the layer to the visualization using \verb|add_layer|, then call \verb|show|:

\begin{minipage}{\textwidth}
\begin{python}
geoplotlib.add_layer(CustomLayer(mydata))
geoplotlib.show()
\end{python}
\end{minipage}

\subsection*{Animation}
A custom layer can be also used for creating animated visualizations. 
Each time the draw method is called, the custom layer can update its state to the next frame. 
As an example, let us imagine having data containing the position of an object over time. 
A simple animation can use a frame counter, and at each frame render only the datapoint at the current instant:

\begin{minipage}{\textwidth}
\begin{python}
class AnimatedLayer(BaseLayer):

  def __init__(self, data):
    self.data = data
    self.frame_counter = 0

  def invalidate(self, proj):
    self.x, self.y = proj.lonlat_to_screen(
      self.data['lon'], self.data['lat'])
    
  def draw(self, proj, mouse_x, mouse_y, ui_manager):
    self.painter = BatchPainter()
    self.painter.points(self.x[self.frame_counter],
                        self.y[self.frame_counter])
    self.painter.batch_draw()
    self.frame_counter += 1
\end{python}
\end{minipage}
Notice that in this case we do not initialize the \verb|BatchPainter| inside \verb|invalidate|, but we create a new one at each frame. 
We also keep track of the current frame with the \verb|frame_counter| variable. 
Even this very simple code is able to visualize a non-trivial animation of an object moving over time.
To produce a movie from the animation, individual frames can be captured using the \verb|screenshot| method, and then combined together.

\subsection*{Colormaps}
Colors can be used as additional mapping for encoding information into a visualization.
Continuous variables (for example points density or the edges distances) can be mapped to a continuous color scale.
The \verb|ColorMap|~class allows a user to perform this conversion.
A \verb|ColorMap| object is constructed by passing any of the matplotlib colorscales, and optionally an alpha value and a number of discretization levels.
The \verb|to_color| method performs the conversion from real value to color:

\begin{minipage}{\textwidth}
\begin{python}
# hot colormap
cmap = ColorMap('hot') 

# Reds colormap with transparency
cmap = ColorMap('Reds', 128) 

# coolwarm colormap with 4 levels
cmap = ColorMap('coolwarm', levels=4) 

# linear scaling
cmap.to_color(10, 100, 'lin')

# logarithmic scaling
cmap.to_color(10, 100, 'log')

# square-root scaling
cmap.to_color(10, 100, 'sqrt')
\end{python}
\end{minipage}
Discrete variables such as categories can be represented using categorical colormaps.
The \verb|colorbrewer| method provides access to the ColorBrewer~\cite{harrower2003colorbrewer} colors.
Categorical colormaps can be also generated from regular colormaps using using \verb|create_set_cmap|:

\begin{minipage}{\textwidth}
\begin{python}
cmap1 = colorbrewer([1,2,3])
cmap2 = create_set_cmap('hsv', [1,2,3])
\end{python}
\end{minipage}

\subsection*{Controlling the map view}
The map view is determined by the projection parameters: the latitude offset, the longitude offset and the zoom level.
By default, the projection is chosen so to fit all selected points, with the maximum zoom level possible.
The view can changed to a specific portion of the map by passing a \verb|BoundingBox| object to the \verb|set_bbox| method.
A \verb|BoundingBox| object defines the map view boundaries, and can be constructed in multiple ways. 
The most direct way is to specify two ranges of latitudes and longitudes. 
Alternatively, a \verb|BoundingBox| can be constructed to fit a subset of points using the \verb|from_points| methods. 

\begin{minipage}{\textwidth}
\begin{python}
bbox1 = BoundingBox(north=51.3, west=-124.3, 
                    south=14.8, east=-56.8)

bbox2 = BoundingBox.from_points(lons, lats)
\end{python}
\end{minipage}

\subsection*{Interactivity}
Finally, geoplotlib allows users to create interactive visualizations by provides support for rendering a user interface, and dynamically changing the visualization on user input:
\begin{itemize}
\item on-screen text such as information or status can be added using the \verb|UiManager| class.
\item mouseover tooltips can be configured on arbitrary graphical elements or screen regions using the \verb|HotspotManager| class.
\item layers can be configured to react to specific key presses by defining a \verb|on_key_release| method
\end{itemize}

\section*{Performance}
\label{sec:performance}
We test the performance of geoplotlib by generating some of the described visualization on a dataset consisting of one million samples, using the default visualization parameters.
All tests consider only the time needed for the actual rendering of the visualization, excluding the time for loading the data.
The measurements are repeated 10 times for each visualization type.
The experiments were performed on a MacBook Pro 2012 with an Intel 2.3 GHz i7 CPU, 8 GB RAM and nVidia GeForce GT 650M GPU.
Table~\ref{tab:performance} shows that in all cases the visualizations require only a few seconds, thus demonstrating that geoplotlib is suitable even for large-scale datasets.

\begin{table}[!h]
\renewcommand{\arraystretch}{1.3}
\caption{Execution time for one million samples}
\label{tab:performance}
\centering
\begin{tabular}{|l|r|r|} 
\hline
\textbf{visualization type} & \textbf{mean time [s]} & \textbf{SD [s]} \\
\hline
dot     &  1.57 & 0.08 \\
\hline
graph   &  1.99 & 0.09 \\
\hline
hist    &  8.12 & 0.55 \\
\hline
kde     &  5.37 & 0.37 \\
\hline
voronoi &  3.08 & 0.66 \\
\hline
\end{tabular}
\end{table}

\section*{Conclusion}
We have presented geoplotlib, a python toolbox for generating geographical visualizations.
We demonstrated how geoplotlib provides a simple yet powerful API to visualize geographical data, greatly facilitating exploratory data analysis of geographical information.
We believe that geoplotlib can become a powerful tool in the data analyst toolbox, both for analyzing complex spatial patterns and for communicating results in forms of geographical visualizations.
Future work includes the addition of more visualization tools, and the integration of spatial analysis methods.

\section*{Acknowledgments}

This work is funded in part by the High Resolution Networks project (The Villum Foundation), as well as Social Fabric (University of Copenhagen).

\end{document}